\documentclass{aastex}
\usepackage{emulateapj5}

\slugcomment{accepted by ApJ}

\shortauthors{Inoue, Hirashita and Kamaya}
\shorttitle{Lyman Continuum Extinction}

\begin{document}

\title{EFFECT OF DUST EXTINCTION ON ESTIMATING STAR FORMATION RATE
OF GALAXIES: LYMAN CONTINUUM EXTINCTION}

\author{AKIO K. INOUE, HIROYUKI HIRASHITA\altaffilmark{1} 
and HIDEYUKI KAMAYA \altaffilmark{2}} 

\affil{Department of Astronomy, Faculty of Science, Kyoto University,
Sakyo-ku, Kyoto 606-8502, JAPAN}
\email{AKI: inoue@kusastro.kyoto-u.ac.jp}

\altaffiltext{1}{Research Fellow of Japan Society for the Promotion of
Science.}
\altaffiltext{2}{Visiting Academics at Department of Physics, Oxford
University, Keble Road, Oxford, OX1, 3RH, UK}

\begin{abstract}

We re-examine the effect of Lyman continuum ($\lambda \leq 912$ \AA) 
extinction (LCE) by dust in H {\sc ii} regions in detail 
and discuss how it affects the estimation of the global star formation 
rate (SFR) of galaxies.
To clarify the first issue, we establish two independent methods for 
estimating a parameter of LCE ($f$), which is defined as the fraction
of Lyman continuum photons contributing to hydrogen ionization in an 
H {\sc ii} region.
One of those methods determines $f$ from the set of Lyman continuum 
flux, electron density and metallicity.
In the framework of this method, 
as the metallicity and/or the Lyman photon flux increase, 
$f$ is found to decrease.
The other method determines $f$ from the ratio of infrared flux
to Lyman continuum flux.
Importantly, we show that $f \la 0.5$ via both methods in many H {\sc ii} 
regions of the Galaxy.
Thus, it establishes that dust in such H {\sc ii} regions absorbs 
significant amount of Lyman continuum photons directly.
To examine the second issue, we approximate $f$ to a function of only 
the dust-to-gas mass ratio (i.e., metallicity), assuming a parameter 
fit for the Galactic H {\sc ii} regions.
We find that a characteristic $\hat{f}$, which is defined as $f$
averaged over a galaxy-wide scale, is 0.3 for the nearby
spiral galaxies.
This relatively small $\hat{f}$ indicates that a typical increment factor 
due to LCE for estimating the global SFR 
($1/\hat{f}$) is large ($\sim 3$) for the nearby spiral galaxies.
Therefore, we conclude that the effect of LCE is not negligible 
relative to other uncertainties of estimating the SFR of galaxies.

\end{abstract}
 

\section{INTRODUCTION}

When we estimate the present star formation rate (SFR) in galaxies, we
use the luminosity of radiation from young massive stars as its indicator.
Various observational quantities are adopted in order to
count the number of photons originating from these stars.
Indeed, we use the data of hydrogen recombination lines,
ultraviolet (UV), infrared (IR), radio, etc., as the indicators of
the photon flux and the SFR (e.g., \citealt{ken98}).
Observational evidence suggests, however, that star-forming regions are
often associated with dust (e.g., \citealt{gla99}, p.\ 125).
Thus, we cannot obtain the true SFR of galaxies unless we
correct observational data for dust extinction.
In other words, understanding the extinction property is important in
estimating the SFR (e.g., \citealt{mad98}).
Therefore, we should examine the properties of dust extinction in
star-forming regions in detail.

The extinction by the interstellar dust from UV ($\lambda > 912$ \AA) to 
near-IR (NIR) has been studied well to date (e.g., interstellar
extinction curve: \citealt{sav79, sea79, cal94, gor00}).
According to those researches, it is widely accepted that 
when we estimate the SFR from the H$\alpha$ luminosity, 
for example, 
we should correct decrement of the line luminosity due to the
UV--NIR extinction.
It can be performed 
by using a proper extinction curve and the observation
of the Balmer decrement (e.g., \citealt{ost89}).
On the other hand, how about the effect of the extinction in the Lyman
continuum band?
Because of the decrement of the stellar radiation owing to the Lyman
continuum extinction (LCE), 
we may underestimate the SFR of galaxies. 
In this paper, this point is extensively examined.

Indeed, many observations of H {\sc ii} regions have revealed that dust
grains are associated with these regions (e.g., \citealt{ish68, har71,
wyn74, fre79, miz82}).
This means a very important fact that the Lyman continuum photons from the
central young massive stars suffer extinction by dust grains within H {\sc
ii} regions.
As a result, the number of Lyman continuum photons contributing to
hydrogen ionization is reduced.
{}From the intensity of recombination lines or thermal radio
radiation, we can only estimate the number of the photons really
contributing to the ionization.
Therefore, we inevitably underestimate the number of Lyman continuum
photons and the SFR unless we correct the observational data for the LCE
by dust.

In studies on each individual H {\sc ii} region,
it has been reported that a part of Lyman continuum photons are
absorbed by dust if H {\sc ii} regions contain dust grains (e.g.,
Petrosian, Silk, \& Field 1972;
\citealt{pan74, nat76, sar77, spi78, mat86, aan89, shi95, bot98}).
According to \cite{pet72}, for instance,
the fraction of Lyman continuum photons used by hydrogen
ionization is estimated to be 0.26 for the Orion nebula.
Their result has shown that the chance which the Lyman continuum
photons are extinguished by dust is significantly large.
On a galactic scale, Smith, Biermann, \& Mezger (1978) have examined LCE
by dust for a large number of radio H {\sc ii} regions in the Galaxy,
and then, have shown about half Lyman continuum photons from exciting
stars in many H {\sc ii} regions are absorbed by dust within these
regions.
However, they did not discuss the effect of the LCE on estimating the
SFR of galaxies.

Unfortunately, it seems that
we often estimate the SFR of galaxies from Lyman continuum
flux (e.g., \citealt{ken83}) without any correction of the LCE effect.
Hence, it is indispensable for us to re-examine LCE by dust within H
{\sc ii} regions and to discuss the effect of LCE on determining
the SFR of galaxies.
First, we formulate the way of estimating the fraction of Lyman continuum
photons used by hydrogen ionization, and determine the fraction in each
individual H {\sc ii} region of the Galaxy in section 2.
We also present another way for estimating LCE by dust in section 3.
Then, we discuss the effect of LCE by dust on estimating the
galaxy-scale SFR in section 4.
Finally, our conclusions are summarized in the last section.

\section{FRACTION OF LYMAN CONTINUUM PHOTONS CONTRIBUTING TO HYDROGEN
 IONIZATION}\label{sec:f_value}

The Lyman continuum photons from young massive stars in an H {\sc ii} 
region are absorbed by dust grains as well as by neutral hydrogen atoms.
Then, the number of Lyman continuum photons contributing to
hydrogen ionization, $N'_{\rm Ly}$, is smaller than the intrinsic number
of Lyman continuum photons, $N_{\rm Ly}$.
We parameterize this effect by $f$, which is defined as the fraction
of Lyman continuum photons contributing to the hydrogen ionization;  
$N'_{\rm Ly} = fN_{\rm Ly}$.
Unfortunately, we are able to estimate only $N'_{\rm Ly}$ from
observations of recombination lines or thermal radio continuum.
This means that the ``real'' SFR, which should be estimated from
$N_{\rm Ly}$, is $1/f$ times larger than the ``apparent'' SFR
estimated from $N'_{\rm Ly}$.
Hence, if the parameter, $f$, is much smaller than unity, the effect of
LCE by dust becomes very important to the estimation of the SFR.
We call $1/f$ {\it correction factor} or {\it increment factor} for the
SFR.
In this section, we examine the value of $f$ in each individual
H {\sc ii} region in the Galaxy to find the correction factor
quantitatively.

\subsection{Formulation}

First, when no dust grains exist in an H {\sc ii} region,
the radius of the region is estimated to be 
the Str\"{o}mgren radius, $r_{\rm S}$. According to \cite{spi78},
we express the number of Lyman continuum photons emitted per unit time,
$N_{\rm Ly}$, by using the Str\"{o}mgren radius, $r_{\rm S}$:
\begin{equation}
 N_{\rm Ly} = \frac{4\pi}{3}{r_{\rm S}}^3 n_{\rm e} n_{\rm p} \alpha^{(2)}\,,
 \label{eq1}
\end{equation}
where $n_{\rm e}$ and $n_{\rm p}$ are the number densities of electrons and
protons, respectively, and $\alpha^{(2)}$ is the recombination
coefficient excluding captures to the $n=1$ level.
Here, we employ the Case B approximation, which is
the assumption that H {\sc ii} regions are optically thick for the
photons of all Lyman-emission lines (e.g., \citealt{ost89}).
We also assume spherical symmetry and spatial uniformity of H {\sc ii}
regions in this paper for simplicity.

Next, we consider the effect of LCE by dust on the size of H {\sc ii}
regions.
The actual radius of the ionized region becomes smaller than the
Str\"{o}mgren radius in equation~(\ref{eq1}) by the LCE effect.
When we express the actual ionized radius as $r_{\rm i}$, we obtain
\begin{equation}
 N'_{\rm Ly} = fN_{\rm Ly} 
 = \frac{4\pi}{3}{r_{\rm i}}^3 n_{\rm e} n_{\rm p} \alpha^{(2)}\,.
 \label{eq2}
\end{equation}
Moreover, if the ratio of $r_{\rm i}$ to $r_{\rm S}$ is expressed by
$y_{\rm i}$, we obtain
the following relation from equations~(\ref{eq1}) and (\ref{eq2}),
\begin{equation}
 f = {y_{\rm i}}^3\,.
 \label{eq3}
\end{equation}

To estimate $f$, we need to find the quantitative relation
between $f$ and the amount of dust.
We determine $f$ in terms of the optical depth of dust for Lyman
continuum photons.
According to \cite{hir01}, then, 
we define $\tau_{\rm S,d}$ as the optical depth of dust for Lyman
continuum photons over the Str\"omgren radius in order to formulate the
dependence of $f$ on the dust-to-gas ratio.
Here, $\tau_{\rm S,d}$ is approximated to be the optical depth at the Lyman
limit (912 \AA).
When we adopt the Galactic extinction curve, the dust extinction at 912
\AA\ is about 13$E_{B-V}$ mag.
Thus, $\tau_{\rm S,d} \simeq 13E_{B-V}/2.5\log e \simeq 12E_{B-V}$.
If we assume that the dust-to-gas mass ratio, $\cal D$, is
proportional to $E_{B-V}/N_{\rm H}$, where $N_{\rm H}$ denotes the
column number density of hydrogen, then, according to \cite{spi78} and
\cite{hir01},
\begin{equation}
 E_{B-V} = \left(\frac{\cal D}{6\times10^{-3}}\right)
 \left(\frac{N_{\rm H}}{5.9\times10^{21}{\rm cm^{-2}}}\right)\,
[\rm mag]\, , 
\end{equation}
where $\cal D$ is estimated with respect to a typical Galactic value,
$6\times10^{-3}$.
Thus, 
\begin{equation}
 \tau_{\rm S,d} = 12 
   \left(\frac{\cal D}{6\times10^{-3}}\right)
   \left(\frac{N_{\rm H}}{5.9\times10^{21}{\rm cm^{-2}}}\right)\,.
 \label{eq19}
\end{equation}

Here, we define $N_{\rm H}$ as the column density over the
Str\"{o}mgren radius.
That is, $N_{\rm H} \equiv n_{\rm H}r_{\rm S}$, where $n_{\rm H}$
denotes the volume number density of hydrogen.
Therefore, equation~(\ref{eq19}) is reduced to 
\begin{equation}
 \tau_{\rm S,d}=0.92\left(\frac{\cal D}{6\times10^{-3}}\right)
   \left(\frac{n_{\rm H}}{10^2\,{\rm cm^{-3}}}\right)^{1/3}
   \left(\frac{N_{\rm Ly}}{10^{48}\,{\rm s^{-1}}}\right)^{1/3}\,,
 \label{eq18}
\end{equation}
where $r_{\rm S}$ is eliminated by equation~(\ref{eq1}), and we suppose 
that $n_{\rm p}\approx n_{\rm e}\approx n_{\rm H}$ (fully ionized
in the ionized region) and that the temperature of H {\sc ii} regions
is $10^4$ K.
Since we cannot observe the real number of Lyman continuum photons,
$N_{\rm Ly}$, we rewrite equation~(\ref{eq18}) in terms of the apparent
one, $N'_{\rm Ly}(={y_{\rm i}}^3N_{\rm Ly})$;
\begin{equation}
 y_{\rm i} \tau_{\rm S,d}=0.92\left(\frac{\cal D}{6\times10^{-3}}\right)
   \left(\frac{n_{\rm H}}{10^2\,{\rm cm^{-3}}}\right)^{1/3}
   \left(\frac{N'_{\rm Ly}}{10^{48}\,{\rm s^{-1}}}\right)^{1/3}\,.
 \label{eq20}
\end{equation}

In addition, \cite{spi78} derived the following relation between
$y_{\rm i}$ and $\tau_{\rm S,d}$:
\begin{equation}
 3\int_0^{y_{\rm i}} y^2 e^{y \tau_{\rm S,d}} dy = 1\,,
 \label{eq22}
\end{equation}
where $y$ is the radius normalized by $r_{\rm S}$.
Integrating equation~(\ref{eq22}), we obtain
\begin{equation}
 {\tau_{\rm S,d}}^3 
 = 3\{e^{\tau_{\rm d}}(\tau_{\rm d}^2 - 2\tau_{\rm d} + 2) - 2\}\, ,
\end{equation}
where $\tau_{\rm d}$ is defined as $y_{\rm i} \tau_{\rm S,d}$.
It is worthwhile to note that $\tau_{\rm d}$ is identified with the
optical depth of dust over the actual ionized radius, $r_{\rm i}$.
According to equation~(\ref{eq3}), $f={y_{\rm i}}^3=(\tau_{\rm
d}/\tau_{\rm S,d})^3$.
Thus, we finally obtain
\begin{equation}
 f = \frac{\tau_{\rm d}^3}
     {3\{e^{\tau_{\rm d}}(\tau_{\rm d}^2 - 2\tau_{\rm d} + 2) - 2\}}\,.
 \label{eq31}
\end{equation}
This is the same as equation~(8) in \cite{pet72}.

Now, once $n_{\rm H}$ ($\approx n_{\rm e}$), $N'_{\rm Ly}$, and $\cal D$ 
are determined, we derive $\tau_{\rm d}(=y_{\rm i} \tau_{\rm S,d})$ via
equation~(\ref{eq20}), and then, we calculate $f$ by using
equation~(\ref{eq31}).
In Figure~1, we show the $f$ vs.\ $\tau_{\rm d}$ relation obtained from
equation~(\ref{eq31}). 
Which quantity determines $f$ the most effectively?
According to figure~1, decrement of $f$ is due to increment of 
the dust optical depth.
Thus, to find the most effective parameter for $f$, let us examine the
parameter dependence of the optical depth.
We find in equation~(7) that the optical depth increases according to 
the dust-to-gas ratio, Lyman continuum flux, and density of H{\sc
ii} regions.
Then, when any of those three parametes increase, $f$ decreases.
By the way, among the power indices in equation~(7),
that of the dust-to-gas ratio is the largest.
Then, we find that $f$ depends most largely on the dust-to-gas ratio. 

One might think that the effect of helium on the ionization
structure of H {\sc ii} regions is important (e.g., \citealt{ost89}). 
In H {\sc ii} regions, there are extremely energetic photons whose
wavelengths are
shorter than 504 \AA. Such photons can ionize neutral helium.
Hence, the number of Lyman continuum photons absorbed by hydrogen seems to
decrease because of the absorption by helium.
However, fortunately, almost all (96\% according to \citealt{mat71})
photons produced by helium recombination can ionize neutral hydrogen.
Therefore, we can assume safely that the number of Lyman continuum photons
does not significantly decrease owing to the presence of helium.
Moreover, the fraction of the photons with shorter wavelengths than 504 
\AA\ is small (0.14 in our calculation adopting the stellar spectra
assumed to be the Plank function and Salpeter's IMF).
Thus, the effect of helium on the photon count is small in estimating
the SFR.
Another effect of helium appears on the electron number density.
For example, in the derivation of equation~(\ref{eq18}), we must replace
$n_{\rm e} = n_{\rm H}$ with $n_{\rm e} = n_{\rm H} + n_{\rm He}$.
Since helium abundance is typically 0.1 by number, 
$n_{\rm e} = 1.1 n_{\rm H}$. 
Hence, we consider this effect to be less significant
than that caused by some other uncertainties, e.g., the change of the
IMF.
Therefore, we neglect these effects of helium in this paper.

\subsection{Value of $f$ for Individual H {\sc ii} Region in the Galaxy}

Let us estimate $f$ for the Galactic H {\sc ii} regions in order to
estimate the effect of LCE by dust on determining the SFR.
To do this, as formulated in the previous subsection,
we need the data set of the number of Lyman continuum
photons, the electron (or hydrogen) number density, and the dust-to-gas
ratio.

The amount of dust is considered to be related to the metallicity.
In this section, we determine the dust-to-gas mass ratio, ${\cal D}$,
from the observational metallicity (abundance of oxygen) for
each H {\sc ii} region, using a global relation between $\cal D$
and (O/H)
proposed and modeled by
Hirashita (1999a,b).\footnote{The parameters are set as
$f_{\rm in,\, O}=0.1$ and $\beta_{\rm acc}=2\beta_{\rm SN}=10$.
Here, $f_{\rm in,\, O}$ is the dust mass fraction in the
material injected from stars, and $\beta_{\rm acc}$ and
$\beta_{\rm SN}$ are defined as gas consumption timescale
(gas mass divided by star formation rate) normalized by dust
growth timescale and by dust destruction timescale, respectively
(see \citealt{hir9a} for details).}
We display in figure~2 the global relation between $\cal D$
and $12+\log({\rm O/H})$.
This is the same as Figure~4 of \cite{hir01}.
Here, we adopt $12+\log({\rm O/H}) = 8.93$ as the solar abundance
(\citealt{cox00}).
We should note here that this $\cal D$--(O/H) relation is the average
relation for the interstellar medium (ISM) in nearby spirals and dwarfs.
That is, we assume that $\cal D$--(O/H) relation of H {\sc ii} regions
is not different significantly from the averaged $\cal D$--(O/H) of ISM.
This assumption may not be very good, but the dust properties in H {\sc
ii} regions are still uncertain.
Hence, as a first step, we apply the $\cal D$--(O/H) relation of the
ISM to each H {\sc ii} region.

We make two sets of the suitable samples to estimate $f$
of each individual H {\sc ii} region.
One is the sample of seven representative giant H {\sc ii} regions in the 
Galaxy.
Their properties are summarized in Table~1.
These H {\sc ii} regions are famous and have been studied well.
Thus, the uncertainties of their data, especially their distance (i.e.,
their luminosity), can be regarded as small.
Their apparent number flux of Lyman continuum photons ($N_{\rm Ly}'$)
are estimated from radio observations and more than about $10^{49}$
s$^{-1}$.
Hereafter, we call this sample {\it luminous} sample.
The determined $f$ values are also shown in column (7) of Table~1.
We find a good agreement between Petrosian et al.'s $f$ and ours for the
Orion nebula ($f=0.26$).
The mean $f$ of this sample is 0.35.

The other sample is a part of H {\sc ii} regions observed and
investigated by \cite{cap00} and \cite{deh00}.
We select the H {\sc ii} regions whose kinematic distance and
photometric distance are consistent within a factor of 1.5 from their
sample.
This is because the uncertainty in distance affects the
estimation of the number of Lyman continuum photons and then prevents us
from determining $f$ reasonably. 
These H {\sc ii} regions tend to be less luminous than the sample of
Table~1 and their apparent number flux of Lyman continuum photons
($N_{\rm Ly}'$) are estimated from their H$\alpha$ luminosities.
Hereafter, the sample is called to {\it less luminous} sample.
Their properties are tabulated in Table~2.
\cite{cap00} also observed the Orion nebula and M16, both of which are
also included in the
sample in Table~1, but their H$\alpha$ luminosities are smaller than
that expected from the radio observations adopted in Table~1.
This is because \cite{cap00} did not observe whole parts of these
nebulae.
Thus, we exclude the Orion nebula and M16 from the sample of Table~2.
The finaly determined $f$ of the sample distributes over 0.5--1,
and the mean is 0.8.

In Figures~3 (a)--(f), 
we present several relations among various parameters of our
sample H {\sc ii} regions in Tables~1 and 2.
Open symbols are H {\sc ii} regions in Table~1, and filled symbols
are those in Table~2.
In the panels of (a)--(c), we show some relations among the basic
parameters to determine $f$. From these panels, we can check
our standpoint for the subsequent considerations.
Firstly, we may find, in panel (a), the sample regions with higher
metallicity have larger observed Lyman continuum flux, $N'_{\rm Ly}$.
Indeed, the sample correlation coefficient is 0.41, so that the
correlation exists with a confidence level of 95\% for the sample.
However, we find no correlation for the sample of Table~2 alone.
We also cannot find the correlation for the extragalactic H {\sc ii}
regions.\footnote{We examine whether H {\sc ii} regions with higher
metallicity have larger luminosity for the data of H {\sc ii} regions in 
M101 presented by \cite{sco92}. The data can be taken from ADC,
NASA/Goddard or CDS, Strasbourg, France via on-line.}
Thus, we consider that the correlation in panel (a) may be caused by our
sample selection.
Hence, we assume it conservatively in this paper that
$N'_{\rm Ly}$ dose not depend on the metallicity. 
In panel (b), the correlation between the metallicity and the
electron number density, $n_{\rm e}$, is not found.
Thus, we dose not consider that $n_{\rm e}$ depends on the metallicity.
In panel (c), interestingly, 
we can find a correlation between $n_{\rm e}$ and
$N'_{\rm Ly}$ for only samples in Table~2.

Next, in the panels of (d)--(f),
we present the dependece of $f$ on the parameters of H {\sc ii} regions.  
The estimated $f$ is regarded as a function of the metallicity in panel
(d), i.e., as the metallicity in an H {\sc ii} region increases, $f$
becomes smaller.
This is because, in our formulation, the dust-to-gas ratio
(i.e., metallicity) has a positive dependence on the optical
depth of dust as shown in equation~(\ref{eq20}), and then, a larger
dust content leads to a larger optical depth of dust and smaller $f$.
Although the correlation in (d) is not the result directly derived
via the analysis of the observational quantities, 
we confirm a trend that $f$ depends on the metalicity
(see also Hirashita et al.\ 2001).
In panel (e), we also find $f$ is related to $N'_{\rm Ly}$ which has
a positive dependence on the dust optical depth (eq.\ [\ref{eq20}]).
This issue will be discussed again in section 3.
We cannot find a significant correlation in panel (f), probably 
because $n_{\rm e}$ has a weak dependence in the dust optical depth
(index 1/3 in eq.\ [\ref{eq20}]) and its dynamic range is small relative
to $N'_{\rm Ly}$ which has the same dependence in equation~(\ref{eq20}).

Now, we concentrate on the dependence of the dust-to-gas
ratio (i.e., metallicity) on $f$, because the variation
of dust-to-gas ratio has the
largest influence on determining the dust optical depth as seen in
equation~(\ref{eq20}).
That is, we assume that the optical depth of dust is a
function of the dust-to-gas ratio only.
If we consider $N'_{\rm Ly}$ and $n_{\rm e}$ dose not depend on $\cal
D$ (or metallicity), we can re-express equation~(\ref{eq20}) as 
\begin{equation}
 \tau_{\rm d} \equiv 
 y_{\rm i}\tau_{\rm S,d} = x \left(\frac{\cal D}{6\times10^{-3}}\right)\,,
 \label{eq30}
\end{equation}
where $\tau_{\rm d}$ is the dust optical depth over the actual ionized
radius (see section 2.1), and $x$ is a factor containing the dependence
of $N'_{\rm Ly}$ and $n_{\rm H} (\approx n_{\rm e})$.
We find that the best fit value of $x$ for the adopted Galactic H
{\sc ii} regions is 1.8, although the dispersion is large.
In Figure~4, we display the estimated $f$ of the individual H {\sc ii}
region in our sample as a function of its dust-to-gas mass ratio,
$\cal D$.
The open and filled symbols represent the sample of Tables~1 and 2,
respectively.
This figure is basically identical to Figure~3 (d).
The solid line is the best fit line ($x=1.8$).
We also show the lines of $x=1.0$, and 3.0 for comparisons.

Here, let us discuss the increment factor for the SFR, $1/f$, which is
introduced at the beginning of section \ref{sec:f_value}.
It is useful for us and readers to check quantitatively how $f$ is
affected by $\cal D$. As a representative case, 
we consider the model of $x=2$, 
which corresponds with $N'_{\rm Ly}\simeq 10^{49} \,{\rm
s^{-1}}$ and $n_{\rm e} \simeq 100\,{\rm cm^{-3}}$ (nearly equal to the
Lyman continuum number flux and electron number density of the Orion
nebula).
Numerically, from the set of equations (7) and (10),
we find $f$ to be about 0.2 for ${\cal D} = 6\times10^{-3}$ (the typical
value of the ISM in the Galaxy; \citealt{spi78}).
If $\cal D$ is 1/3, 1/10, or 1/100 of the Galactic value, $f$
becomes 0.6, 0.9, or nearly unity, respectively.
Thus, it is found that the increment factor for the SFR, 
$1/f$, increases from 
almost unity to about 5 as $\cal D$ increases from $6\times10^{-4}$
(1/10 of a typical Galactic value) to $6\times10^{-3}$.
Clearly, the correction factor for the SFR is very sensitive to the
dust-to-gas ratio (or metallicity), especially around ${\cal D} \sim
10^{-3}$.
Hence, we should determine $\cal D$ precisely
to find an actual $f$ and SFR.
In any case, we must take account of the effect of LCE by dust at
least for the objects whose metallicity is as high as that
of the Orion nebula.

\cite{smi78} have also determined $f$ for a numerous
radio sample of giant H {\sc ii} regions in the Galaxy.
Although our equation~(\ref{eq20}) is equivalent to their
equation~(A.2), there is a difference in the method of determining the
dust-to-gas ratio of H {\sc ii} regions. Indeed, they used an empirical
relation of the absorption cross section as a function of the
Galactocentric radius (\citealt{chu78}), whereas we determine $\cal D$
for each H {\sc ii} region individually as described above.
The determined mean $f$ for the giant H {\sc ii} regions by \cite{smi78} 
is 0.56.
Since $N'_{\rm Ly}$ of their sample is larger than
$10^{49}~{\rm s^{-1}}$,
their sample H {\sc ii} regions correspond to our sample in Table~1,
which has mean $f=0.35$.
The estimated $f$ values by us may be systematically small.
This is due to the differences of some adopted parameters in
calculation, for example, $N_{\rm H}/E_{B-V}$.
Also, \cite{smi78} took account of the effect of helium which we
neglect as mentioned in the previous subsection.
The parameter of $f$ in Smith et al. is thus determined 
as the sum of the fractions of Lyman continuum photons absorbed
by both hydrogen and helium.
This increment by helium in $f$ is about several \% (up to 10\%) in
their framework.
If the increment is removed, Smith's mean $f$ will approach our value.

Throughout this section, we have found that the determination of
$\cal D$ is very important to estimate the ``real'' SFR (or $f$) from
an observational count of Lyman continuum photons.  
To obtain $\cal D$ accurately, it is a recommendable way
that we use the luminosity of IR thermal emission of dust in
H {\sc ii} regions.
Since dust grains absorb the radiative energy from the young
massive stars and then re-emit the energy in the IR range,
the amount of dust is determined from the IR luminosity.
In the next section, we discuss LCE by dust from observation of IR
luminosity.

\section{CORRELATION BETWEEN IR AND H$\alpha$/RADIO LUMINOSITIES
FOR H {\sc ii} REGIONS}

In the previous section, we estimate the amount of dust
(the dust-to-gas ratio) in H {\sc ii} regions from their metallicity.
Since the amount of dust is also reflected in the IR emission, $f$ is
also determined from IR luminosity independently.
In this section, then, 
we determine $f$ by using the IR luminosity of H {\sc ii} regions
in order to check consistency between the results of the previous and
the current section.

Based on the theory of the IR emission from dust in H {\sc ii} regions
by \cite{pet72}, we formulate the relation between the observed IR
luminosity, $L^{\rm dust}_{\rm IR}$, and the observed number of Lyman
continuum photons, $N'_{\rm Ly}$ (Inoue, Hirashita, \& Kamaya 2000, 2001).
The derived formula is 
\begin{equation}
 \frac{L^{\rm dust}_{\rm IR}(8-1000{\rm \mu m})/L_\sun}
      {N'_{\rm Ly}/5.63\times10^{43}~{\rm s^{-1}}}
 = \frac{0.44-0.28f+0.56\epsilon}{f}\,,
  \label{eq40}
\end{equation}
where $\epsilon$ is the average efficiency of the dust absorption for
UV--optical photons from OB stars.
Here, we have also adopted the Salpeter's IMF (the stellar mass range is
0.1--100 $M_\sun$) and proper stellar properties.
Moreover, the intrinsic number of Lyman continuum photons, $N_{\rm Ly}$, 
is $5.63\times10^{43}~{\rm s^{-1}}$ per unit solar luminosity of the
bolometric luminosity of OB stars, in our calculation.
As defined in the previous section, $N'_{\rm Ly}=fN_{\rm Ly}$.
Instead of determining the SED of the IR radiation, we consider the
total IR luminosity of dust, $L^{\rm dust}_{\rm IR}$, in the range of
8 -- 1000 $\mu$m which covers almost the whole wavelength range of dust
emission.

First, let us consider the correlation between the luminosities of IR
and H$\alpha$ for H {\sc ii} regions.
We make a cross-reference between the sample of \cite{cap00} and
the {\it IRAS} Point Source Catalog ({\it IRAS} PSC; \citealt{joi85}).
We select the sample H {\sc ii} regions by the following criteria;
(a) the separation between the positions of the H$\alpha$ observations
and sources in {\it IRAS} PSC is within $100''$.
(b) $\log{(F_{25}/F_{12})} \geq 0.4$.
(c) $\log{(F_{60}/F_{25})} \geq 0.25$.
(d) $F_{100} \geq 80$ Jy.
Here, $F_{12}$, $F_{25}$, $F_{60}$, and $F_{100}$ are the {\it IRAS} fluxes at
12, 25, 60, and 100 $\mu$m, respectively.
The criterion (a) is based on the typical radius of optical H {\sc ii}
regions ($\sim$ 1--10 pc) and {\it IRAS} positional error ($\la 20''$,
\citealt{bei85}).
The criteria (b) -- (d) are proposed by \cite{hug89}.
The confidence level of the set of criteria for true association between
{\it IRAS} sources and H {\sc ii} regions is more than 80\%
(\citealt{hug89, cod94}).
As a result, we obtain 18 H {\sc ii} regions (10 samples are also in
Table~2) and their properties are found in Table~3.

When we suppose that the dust emission can be fitted by the
modified black-body radiation of 30 K (spectral index is 1), {\it IRAS}
40 -- 120 $\mu$m luminosity, $L_{IRAS}$, is
$L_{IRAS}(40-120~{\rm \mu m}) = 0.6 L^{\rm dust}_{\rm IR}(8 - 1000~
{\rm \mu m})$.
This correction factor is consistent with that of \cite{cal00}.
On the other hand, $N'_{\rm Ly}$ is estimated in terms of the
H$\alpha$ luminosity corrected 
for interstellar extinction:  $N'_{\rm Ly}/{\rm s^{-1}} =
2.83\times10^{45} L_{\rm H\alpha}/L_\sun$, where we assume the Case B
and the electron temperature of $10^4$ K (\citealt{ost89}).
Thus, equation~(\ref{eq40}) is reduced to
\begin{equation}
 \frac{L_{IRAS}}{L_{\rm H\alpha}} 
 = \frac{F_{IRAS}}{F_{\rm H\alpha}}
 = 31\left(\frac{0.44-0.28f+0.56\epsilon}{f}\right)\,,
 \label{eq41}
\end{equation}
where $F_{IRAS}$ and $F_{\rm H\alpha}$ are fluxes of {\it IRAS} and
H$\alpha$, respectively.
This is a theoretical relation between $F_{IRAS}$ and $F_{\rm
H\alpha}$.
Since a flux ratio does not depend on the distance to the
objects, equation~(\ref{eq41}) is free from the uncertainty in the 
distance.

The average efficiency of dust absorption for UV--optical photons from OB
stars, $\epsilon$, is defined as $\epsilon \equiv 1 - 10^{-0.4 \langle
A_\lambda \rangle}$, where $\langle A_\lambda \rangle$ is an average
dust absorption over the wavelength range of
$\lambda=1000$ -- 4000 \AA\ in units of magnitude.
This range is suitable for our purpose because the number of
photons of $\lambda >$ 4000 \AA\ emitted by OB stars are negligible.
If we adopt the Galactic extinction curve (\citealt{sav79}),
then we have $\langle A_\lambda \rangle \simeq 7.2 E(B-V)$.
Thus, we obtain the following equation;
\begin{equation}
 \epsilon = 1 - 10^{-3E(B-V)}\,.
 \label{eq42}
\end{equation}
Using this equation, we find $\epsilon \sim 1$ for the H {\sc ii} 
regions in Table~3.

Once we obtain the value of $\epsilon$ for each H {\sc ii}
region from equation (\ref{eq42}), we can determine each $f$ from IR
and H$\alpha$ fluxes by using equation~(\ref{eq41}).
We show the estimated $f$ in column (7) of Table~3.
By definition, $f \leq 1$.
However, for some sample H {\sc ii} regions, $f$ appears to be greater
than unity.
The reason for the discrepancy will be discussed in the last 
paragraph of this section.

By the way, we compare the theoretical $F_{IRAS}$--$F_{\rm H\alpha}$
relation with observational data on the $F_{IRAS}$ vs.\
$F_{\rm H\alpha}$ plane in Figure~5.
Adopting $\epsilon=1$, we make three lines for three values of $f$; the
solid, dotted, and dashed lines for $f=1$, 0.5, and 0.1,
respectively.
Moreover, we make the dash-dotted line for the set of $f=1$,
$\epsilon=0$.
This corresponds to the case that only the Lyman $\alpha$
photons are absorbed by dust grains.
As long as all Lyman $\alpha$ photons are absorbed by dust
within H {\sc ii} regions and re-emitted in the IR range
(\citealt{spi78, hir01}), 
the area below the dash-dotted line on the $F_{\rm
IRAS}$ vs.\ $F_{\rm H\alpha}$ plane is the forbbiden area
within the framework of our theory.
That is, the dash-dotted line stands for the lower boundary of {\it
IRAS} flux corresponding to H$\alpha$ (Lyman continuum photon) flux in
our model.
While the correlation between $F_{IRAS}$ and $F_{\rm
H\alpha}$ is expected theoretically from equation~(\ref{eq41}), we do
not find a significant correlation in observational data points in
figure~5 (the sample correlation coefficient is 0.35).\footnote{If
the correlation coefficient is more than 0.45 for 18 samples,
we can conclude that a correlation exists with a confidence level of
95\%.}
This point is also examined in the final part of this section.

On the other hand, it is well known that there is a tight observational
correlation between the IR luminosity and the number of Lyman continuum
photons estimated from the radio luminosity (\citealt{wyn74}).
Using equation~(\ref{eq40}), we compare this observational
correlation with the theoretical one in figure~6.
Here, we read the data directly from the figure in \cite{wyn74}
and re-plotted them.
The sample correlation coefficient of their data is 0.96, showing
a stong correlation.
The parameters of lines are the same as those in figure~5. 
Comparing the data and the model lines, 
we consider the IR luminosity of \cite{wyn74} (40 -- 
350 $\mu$m) is nearly equal to the whole dust-IR luminosity in
equation~(\ref{eq40}) because the dust luminosity of the wavelength
range of $\lambda<40~\mu{\rm m}$ or $\lambda>350~\mu{\rm m}$ is
negligible (\citealt{cal00}).
{}From figure~6, we find that the theoretical lines agree with the data
of \cite{wyn74} excellently.
This shows that our model is supported by the observation.

If we assume $\epsilon=1$ for the sample of \cite{wyn74}, we can
determine their $f$ from equation~(\ref{eq40}).
We find $f=0.45$ as an average in the current sample.
Now, we compare it with another mean $f$ value estimated 
by the method of the previous section.
Since H {\sc ii} regions in figure~6 have $N'_{\rm Ly} \ga 10^{48}$
s$^{-1}$, 
we shall choose 18 regions whose $N'_{\rm Ly}$ is greater than
$10^{48}$ s$^{-1}$ from Tables 1 and 2.
Thus, we obtain a mean $f$ as being 0.56 by the method of section 2.
Fortunately, we find a good agreement between the two $f$'s.

By the way, 
$f$ values estimated by the both methods are not fully consistent
with each other. 
For example, the sample of \cite{wyn74} includes the Orion nebula, M8,
and M17 as the sample of Table~1 does.
Their $f$ values via equation~(\ref{eq40}) are 0.44, 0.59, and 0.59,
respectively.
Obviously, these values are larger than those of the Orion nebula, M8, 
and M17 in Table 1.
This may be caused by the aparture difference between two photometories,
IR and radio (see also the last paragraph in this section).
This discrepancy may also result from our assumption  in
equation~(\ref{eq40}) that all of the
energy of photons not
used by hydrogen ionization is converted into IR radiation.
Generally, the efficiency of conversion to IR radiation is never unity
because of the escape of Lyman continuum photons from H {\sc ii} regions
or the scattering of these photons by dust.
As a result, we expect rather large IR luminosities for the H {\sc ii}
regions by using equation~(\ref{eq40}).

Moreover, 
we cannot find in Figure~6 the trend that the more luminous H {\sc ii}
regions have the smaller $f$. This trend was seen in section 2.1,
where we determined $f$ from the metallicity.
This may be the effect of IR cirrus.
The observed IR flux originates from dust associated
with the star forming regions and from diffuse dust heated by the
interstellar radiation field.
If we consider that the IR surface brightness of the cirrus component
is almost constant in a galaxy, the effect of cirrus contamination in
the observed IR flux of an H {\sc ii} region is larger as its IR flux
is lower.
In the case, we may overestimate the IR luminosity for low luminosity
regions, so that we may underestimate $f$ of such regions.

Figure~6 has other interesting implications: 
First, the dash-dotted line ($f=1$ and $\epsilon =0$) means that
only Lyman $\alpha$
photons heat dust grains and only this energy is re-emitted in IR.
Obviously the predicted IR luminosity in this case falls short of
the observed one.
Next, the solid line ($f=1$ and $\epsilon =1$) represents the case
that only the energy of
photons whose wavelength is longer than the Lyman limit
is absorbed by dust and re-emitted in IR.
Even in this case, the predicted IR luminosity is not
enough to explain the observed one.
Hence, we conclude that a part of Lyman continuum photons ought to be
aborbed and reprocessed by dust.
The fraction of Lyman continuum photons absorbed by dust is about 0.5 or
more.
In terms of the fraction of the photons used by hydrogen ionization, 
$f \la 0.5$ in many H {\sc ii} regions of our Galaxy.
This is one of the main conclusions in the current paper.

To close this section, we must discuss a few remaining questions 
in the previous paragraphs:
Why is $f$ determined by our theory greater than unity for some H {\sc
ii} regions in Table~3?
Why cannot we find the observational correlation between $F_{IRAS}$
and $F_{\rm H\alpha}$ in figure~5?
These may be because we use the {\it IRAS} PSC.
Since the Galactic H {\sc ii} regions lie close to us and extend over
large areas, they can never be regarded as point-like sources.
Hence, the IR flux for H {\sc ii} regions in {\it IRAS} PSC 
may be underestimated 
since the aperture effect of H {\sc ii} regions is not neglected.
Therefore, {\it IRAS} PSC may not be suitable when we measure the IR
flux of Galactic H {\sc ii} regions (see also \citealt{cha95}).
Conversely, the correlation between the IR luminosity and the number of
Lyman continuum photons inferred from radio observations is tight.
This observational correlation is explained very well by our theory.
This is because the IR luminosities of sample in \cite{wyn74} are
observed in detail individually.
Seeing the observational tight correlation in figure~6, the absence
of the correlation in figure~5 is caused by flux underestimation of {\it
IRAS} PSC due to the finite aperture effect.
Thus, if we use the {\it image} data of the H {\sc ii} regions observed
by {\it IRAS} or recent/future facilities such as {\it ISO}, SOFIA, {\it
SIRTF}, or ASTRO-F (see \citealt{tak99} for a summary of the facilities),
the observational correlation between $F_{\rm IR}$ and $F_{\rm
H\alpha}$ will appear.

\section{EFFECT OF $f$ ON GALACTIC SCALE}

In this section, we estimate $f$ averaged on galactic scale for 
the nearby spiral galaxies.
Then we discuss the effect of LCE by dust on determining their
galaxy-wide SFR.
In the privious sections, we have discussed LCE by dust in H {\sc ii}
regions.
As shown in section 2, the parameter $f$ strongly
depends on the dust-to-gas mass ratio.
For the Galactic H {\sc ii} regions, we show that $x\sim2$ is
appropriate in equation~(\ref{eq30}).
That is, the Orion nebula is likely to be a typical Galactic H {\sc ii}
region in gas density and Lyman continuum flux.
Here, we consider the Orion nebula is also a typical H {\sc ii} region
in the other galaxies.
In short, we assume $x=2$ in equation~(\ref{eq30}) to be universal 
for the galaxies.
Then, we estimate a typical $f$ for the nearby spiral galaxies from
their dust-to-gas ratio.
By the way, we can determine $f$ of the galaxies from
equation~(\ref{eq41}) as perfomed for the Galactic H {\sc ii}
regions in section 3.
This alternative way is discussed in the last part of this section.

In order to determine their dust-to-gas ratio,
we estimate the amount of dust mass in the nearby spiral galaxies from
their IR emission re-processed by dust.
As considered in the final paragraph of section 3,
we do not use {\it IRAS} PSC since we want to estimate the entire IR
luminosity of the sample galaxies safely.
We adopt only the sample galaxies whose entire images by {\it
IRAS} are available.
We note another issue here.
If we convert the IR emission calculated only from {\it IRAS} data
to the dust mass, the total amount of dust will be underestimated. 
This is because there is large amount of cold dust (10--20 K) in the
spiral galaxies, even in those with starburst regions (\citealt{cal00}
and references therein).
Such cold dust cannot be traced by {\it IRAS} bands.
On the other hand, {\it ISO} has better detectability for the cold dust
than {\it IRAS} because {\it ISO} covers longer wavelengths than {\it
IRAS}.
Thus, we expect that {\it ISO} data improve the estimation of
dust mass.

\cite{alt98} present the imaging properties of seven nearby spiral
galaxies obtained by both {\it IRAS} and {\it ISO}.
Their sample galaxies are 
NGC 134, NGC 628 (M74), NGC 660, NGC 5194 (M51),
NGC 5236 (M83), NGC 6946, and NGC 7331.
Some basic properties of these galaxies are tabulated in tables~1, 2 and 
3 of \cite{alt98}.
We estimate the dust-to-gas mass ratio for Alton's sample galaxies, and
then, determine their $f$ averaged over galactic scale. 
To avoid the confusion from $f$ of the H {\sc ii} regions,
in this section, we express the galactic $f$ as being $\hat{f}$.
Of course, each sample galaxy has unique $\hat{f}$. 

We, first, calculate the dust temperature of the sample galaxies
from the ratio of flux densities at 100 and 200 $\mu$m.
In this calculation, we assume the spectrum of dust emission to be the
modified black-body radiation with the spectral index of 1.
Here, we note that the determination of the dust temperature strongly
affects the successive estimation of the dust mass.
The obtained dust temperatures are shown in column (5) of Table 4.
Next, we estimate the dust mass, $M_{\rm d}$, in the sample galaxies
from the observed flux density at 100 $\mu$m via the following equation:
\begin{equation}
 M_{\rm d} = \frac{4}{3} a\rho D^2 \frac{F_\nu}{Q_\nu B_{\nu}(T)}\,,
 \label{eq50}
\end{equation}
where $a$, $\rho$, and $D$ are the grain radius, grain density, and
galaxy distance, respectively.
$F_\nu$, $Q_\nu$, and $B_{\nu}(T)$ are the observed flux density, grain
emissivity, and the Plank function of the temperature $T$ at the
frequency $\nu$, respectively.
According to \cite{hil83}, we set $(4/3)a\rho/Q_\nu\simeq0.04$ g
cm$^{-2}$ at the frequency of 100 $\mu$m ($a=0.1$ $\mu$m and $\rho=3$ g
cm$^{-3}$).
The estimated dust masses are tabulated in column (6) of Table~4.
Then, we determine the dust-to-gas mass ratio, $\cal D$, of the sample
galaxies by using their total (H {\sc i} + H$_2$) gas masses reported in 
literatures (column 7).
The values of $\cal D$ are in column (8), and the average is about
$5\times10^{-3}$.
This is quite consistent with $\cal D$ estimated by \cite{alt98}.
\cite{sti00} also claim that the median value of the dust-to-gas mass
ratios of sample galaxies of ISOPHOT 170 $\mu$m Serendipity Survey
(\citealt{bog96}) is about $4\times10^{-3}$.
Thus, our estimation of $\cal D$ is reasonable.

It still remains open whether $\cal D$ in H {\sc ii} regions can be
approximated to an averaged $\cal D$ of the ISM in the galaxies.
As seen in subsection 2.2, there is a large variation of $\cal D$ (i.e.
observed metallicity) of the Galactic H {\sc ii} regions.
Moreover, most of the sample H {\sc ii} regions in Tables~1 and 2 have
smaller $\cal D$ than the typical value for the Galactic ISM
($6\times10^{-3}$; Spitzer 1978).
It may suggest that $\cal D$ in H {\sc ii} regions is systematically
smaller than that of the ISM.
However, it is reported that the mean $\cal D$ in the ISM is not
so different from that in H {\sc ii} gas.
\cite{sod97} have determined the dust-to-gas mass ratio of the
H {\sc i}, H$_2$, and low-density
($n_{\rm e} \sim 10\, {\rm cm}^{-3}$) H {\sc ii}
gases in the Galaxy from the data of $COBE$.
According to them, the mean value of $\cal D$ in the Galactic ISM (H
{\sc i} + H$_2$ + H {\sc ii}) is 4.6--6.6$\times10^{-3}$, and that in H
{\sc ii} gas is 5.1--10$\times10^{-3}$.
With the uncertainty of a factor of 2, therefore, we assume that
$\cal D$ in H {\sc ii}
regions is approximated to the mean $\cal D$ of the ISM in the
galaxies.

Here, let us calculate a characteristic value of the $\hat{f}$-parameters 
for the sample seven spiral galaxies, adopting their dust-to-gas mass
ratio (column 8 in Table 4), equations~(\ref{eq31}) and
(\ref{eq30}) with $x=2$.
The derived values of $\hat{f}$ are tabulated in column (9) of Table 4.
The averaged $\hat{f}$ is 0.3.
This is nearly equal to the mean $f$ of the sample H {\sc ii}
regions in Table~1 which are more luminous than the Orion nebula.
This small $\hat{f}$ means that the correction factor of the SFR
($1/\hat{f}$; see also section \ref{sec:f_value}) for these
spiral galaxies is large (typically $\sim 3$).

Although $f$ values of H {\sc ii} regions
distribute over a wide range (0.02--1) as shown in Tables~1 and 2, 
$\hat{f}$ is approximately equal to the averaged value of the luminous
H {\sc ii} regions (i.e., the sample in Table 1).
This may be caused by the effect of luminosity function of H {\sc ii}
regions.
Indeed, it is implied that the total luminosity of H {\sc ii} regions
is dominated by the regions more luminous than the Orion nebula
(\citealt{ken89, wal92, wyd97}).
Thus, $\hat{f}$  may be shifted toward smaller values because we expect
smaller $f$ in more luminous H {\sc ii} regions (see section 2.1).
Therefore, we think it reasonable that a characteristic $\hat{f}$ for nearby
spiral galaxies is about 0.3.

Now we discuss the importance of the increment of a factor of 
$1/\hat{f} \sim 3$ for the estimated SFR.
There are two other major uncertainties for the estimation of the SFR.
One is related to the initial mass function (IMF).
The choices of a specific IMF and its upper/lower limit masses affect the
coefficient of the conversion formula of the SFR.
The uncertainties are within a factor of about two (e.g.,
\citealt{ino00}).
The other uncertainty is related to stellar properties.
The choices of the stellar properties adopted affect the conversion
formula of the SFR.
\cite{ken94} derived the formula, ${\rm
SFR}/M_\sun~{\rm yr^{-1}}=3.05\times10^{-8}L_{\rm H\alpha}/L_\sun$,
adopting their stellar population synthesis model and Salpeter IMF
(0.1--100$M_\sun$).
We also derive the same type formula by using the same IMF
but by using very simplified stellar properties (\citealt{ino01}).
In comparison with them, we find that the uncertainty due to the adopted
stellar properties is at most a factor of two.
Therefore, we conclude that the effect of LCE by dust may be more
significant than the uncertainties due to the specific IMF or the
stellar properties for nearby spiral galaxies.

Finally, we discuss the alternative way for determining galaxy-wide
$\hat{f}$.
We can use equation (\ref{eq41}) to do it.
However, there are three difficulties to determine $\hat{f}$ reasonably.
(i) 
One is the uncertainty of the cirrus fraction of the IR luminosity of
the galaxies.
It is widely believed that the IR luminosity of the galaxies consists of
two components: the warm component originating from the dust nearby the
star forming regions and the cirrus component originating from the
diffuse cold dust heated by the interstellar radiation field from the
older stars (e.g., \citealt{lon87}).
It is a very difficult task to subtract the cirrus component from the
observed IR luminosity of the galaxies in order to obtain the ``pure''
IR luminosity originating from the star forming regions.\footnote{We
assumed implicitly in the previous sections that the IR emission from
the H {\sc ii} regions is not contaminated significantly by the cirrus
emission. This is because the H {\sc ii} regions are the star forming
regions themselves. However, we may not be able to neglect the
contamination of cirrus for H {\sc ii} regions with low luminosity.}
Since the cirrus fraction differs from galaxy to galaxy
(e.g., 0.3--0.8; \citealt{lon87}), the subtraction of cirrus
causes a significant uncertainty in the estimation of $\hat{f}$.

The other difficulties are the correction of the H$\alpha$ luminosity
for (ii) the interstellar extinction and (iii) the [N {\sc ii}]
contamination.
Indeed, \citet{ken83} noted that the interstellar extinction at the
wavelength of H$\alpha$ is 1.1 mag on average for his sample galaxies.
That is, the increment correction factor for H$\alpha$ luminosity is
2.8.
For that case, we do not neglect such large extinction in order to
determine $\hat{f}$ reasonably.
At that time, we may also have to correct 
the inclination of the galaxies along the line of 
sight to find the correct $\hat{f}$. 
The [N {\sc ii}] contamination is more complicated if we do not have
sufficient resolution of the spectrograph.
Fortunately, the H$\alpha$ flux of \cite{cap00} is not
contaminated by [N {\sc ii}] because of their high dispersion of 
spectrograph.
Then, our main conclusions on $f$ of the Galactic H {\sc ii} regions are
not altered.
However, we are sure of that the [N {\sc ii}] line contaminates the
galactic H$\alpha$ emission (e.g. \citealt{ken83}).
When we use equation (\ref{eq41}) to find $\hat{f}$, therefore, 
we always need the H$\alpha$ luminosity corrected for the interstellar
reddening and [N {\sc ii}] contamination to find the true $\hat{f}$.
Unfortunately, we face the deficiency of the data to correct the
H$\alpha$ luminosity of each individual galaxy.
Because of several large uncertainties, thus,
we cannot safely determine $\hat{f}$ of the galaxies via
equation~(\ref{eq41}) in this paper.
This problem is examined by us in the near future.

\section{CONCLUSIONS}

For each individual H {\sc ii} region in the Galaxy, we examine LCE by
dust within such regions.
Then, we discuss the LCE effect on determining the SFR of galaxies.
We reach the following conclusions.

[1] We estimate the fraction of Lyman continuum photons contributing to
hydrogen ionization within the Galactic H {\sc ii} regions, $f$.
The estimated $f$ values distribute over a wide range (0.02--1).
In our framework, an H {\sc ii} region with higher metallicity and
larger Lyman photon flux tends to have a smaller $f$.

[2] From the correlation tests in Figure 3, we
find that $f$ is approximately treated as a function of only the
dust-to-gas mass ratio, $\cal D$.
For our sample of the Galactic H {\sc ii} regions, the Orion nebula is
regarded as a typical region in gas density and Lyman continuum flux.

[3] We explain theoretically the observed correlation between IR
luminosity and radio luminosity of the Galactic H {\sc ii} regions.
Then, we manifest that a part of Lyman continuum photons (about a half or
more) ought to be absorbed by dust directly in many regions.

[4] We do not find the observational correlation between fluxes of {\it
IRAS} PSC and H$\alpha$ for our sample H {\sc ii} regions, which is
expected by our model theoretically.
This may indicate that {\it IRAS} PSC is not suitable to determine the IR
flux of nearby extended sources like Galactic H {\sc ii} regions as
pointed out in literatures.

[5] If we assume that $\cal D$ in H {\sc ii} regions is not
significantly different from a mean $\cal D$ of the ISM, we
estimate $f$ averaged over galactic scale to be about 0.3 for the nearby
spiral galaxies (i.e. $\hat{f} \sim 0.3$).
Consequently, the increment factor for the SFR due to the LCE by dust
($1/\hat{f}$) is about 3.
The correction factor is larger than the uncertainties of determining
the SFR due to the choice of a specific IMF or stellar properties
(within a factor of 2).
Therefore, we cannot neglect the effect of LCE by dust when we estimate the
SFR in galaxies.

\acknowledgments

We thank the anonymous referee for very helpful comments, which improved 
quality of the paper significantly.
We are also grateful to Tsutomu T. Takeuchi for careful reading
of this paper and giving us very useful comments.
One of us (H.H.) acknowledges the Research Fellowship of
the Japan Society for the Promotion of Science for Young
Scientists.
We have made extensive use of NASA's Astrophysics Data System Abstract
Service (ADS).

\clearpage

\begin{deluxetable}{lccccccc}
\footnotesize
\tablecaption{Representative Giant Galactic H {\sc ii} Regions}
\tablewidth{0pt}
\tablehead{\colhead{Object} & \colhead{Distance} & \colhead{$N'_{\rm Ly}$} 
& \colhead{$n_{\rm e}$} & \colhead{$12+\log{\rm (O/H)}$}
& \colhead{$\cal D$} & \colhead{$f$} & \colhead{Reference} \\
\colhead{} & \colhead{(kpc)} & \colhead{($10^{48} {\rm s}^{-1}$)} 
& \colhead{(${\rm cm^{-3}}$)} & \colhead{} 
& \colhead{($10^{-3}$)} & \colhead{}\\ 
\colhead{(1)} & \colhead{(2)} & \colhead{(3)} & \colhead{(4)} 
& \colhead{(5)} & \colhead{(6)} & \colhead{(7)} & \colhead{(8)}\\}

\startdata
Orion   & 0.5 & 7.59 & 150 & 8.76 & 5.0 & 0.26 & 1 \\
M8      & 1.1 & 10.6 & 60 & 8.74 & 4.5 & 0.37 & 1 \\
M16     & 2.5 & 29.1 & (100) & 8.76 & 5.0 & 0.16 & 1 \\
Rosette & 1.4 & 43.2 & 13 & 8.20 & 0.79 & 0.85 & 2 \\
M17     & 1.8 & 143  & (100) & 8.81 & 6.3 & 0.016 & 1 \\
Carina  & 1.4 & 167  & 13 & 8.49 & 1.8 & 0.56 & 2 \\
NGC3603 & 8.5 & 897  & 25 & 8.51 & 2.0 & 0.23 & 2 \\
\enddata

\tablecomments{Col.(3): Number of Lyman continuum photons 
 estimated from radio observations by using equation~(2) in
 \cite{mez74}. Col.(4): Averaged electron number density from
 \cite{ken84}. Two parentheses are assumed values. Col.(5): Observed
 Oxygen abundance from \cite{ken00} and references therein. Col.(6):
 Determined dust-to-gas mass ratio by using the $\cal D$--(O/H)
 relation of figure~2 (see also \citealt{hir9a,hir9b}). Col.(7): Estimated
 $f$ parameter by using equations~(7), and (10).}

\tablerefs{Sources of radio observations: 1. \cite{mez67}; 2. \cite{sha70}}

\end{deluxetable}

\clearpage

\begin{deluxetable}{lcccccc}
\footnotesize
\tablecaption{Galactic H {\sc ii} Regions of \cite{cap00}}
\tablewidth{0pt}
\tablehead{\colhead{Object} & \colhead{$L_{\rm H\alpha}$} 
& \colhead{$N'_{\rm Ly}$} & \colhead{$n_{\rm e}$} 
& \colhead{$12+\log{\rm (O/H)}$} & \colhead{$\cal D$} & \colhead{$f$}\\
\colhead{} & \colhead{($10^{35}$ erg s$^{-1}$)} 
& \colhead{($10^{48} {\rm s}^{-1}$)} & \colhead{(${\rm cm^{-3}}$)} 
& \colhead{} & \colhead{($10^{-3}$)} & \colhead{}\\ 
\colhead{(1)} & \colhead{(2)} & \colhead{(3)} & \colhead{(4)} &
 \colhead{(5)} & \colhead{(6)} & \colhead{(7)}\\}

\startdata
S54     & 42.0 & 3.08 & 245  & 8.60 & 2.7  & 0.54 \\
S88     & 2.17 & 0.16 & 28   & 8.43 & 1.5  & 0.94 \\
S93     & 8.65 & 0.63 & 269  & 8.57 & 2.5  & 0.71 \\
S101    & 3.61 & 0.26 & 7    & 8.34 & 1.1  & 0.97 \\
S104    & 76.7 & 5.62 & 56   & 8.51 & 2.0  & 0.71 \\
S127    & 38.4 & 2.81 & 545  & 8.23 & 0.79 & 0.80 \\
S131    & 0.20 & 0.01 & 18   & 8.39 & 1.3  & 0.98 \\
S142    & 16.6 & 1.22 & 21   & 8.27 & 0.89 & 0.94 \\
S148    & 58.2 & 4.27 & 235  & 8.30 & 1.0  & 0.78 \\
S153    & 8.90 & 0.65 & 106  & 8.12 & 0.59 & 0.94 \\
S156    & 59.4 & 4.35 & 907  & 8.42 & 1.5  & 0.55 \\
S168    & 18.4 & 1.35 & 137  & 8.30 & 1.0  & 0.87 \\
S184    & 4.56 & 0.33 & 90   & 8.39 & 1.4  & 0.90 \\
S206    & 31.8 & 2.33 & 313  & 8.37 & 1.3  & 0.75 \\
S209    & 201  & 14.7 & 645  & 8.18 & 0.71 & 0.69 \\
S211    & 26.0 & 1.91 & 135  & 8.03 & 0.48 & 0.93 \\
S219    & 4.26 & 0.31 & 166  & 8.27 & 0.89 & 0.92 \\
S252    & 5.22 & 0.38 & 144  & 8.35 & 1.2  & 0.89 \\
S257    & 2.55 & 0.19 & 160  & 8.17 & 0.69 & 0.95 \\
\enddata

\tablecomments{Col.(2): H$\alpha$ luminosities are calculated by 
 $L_{\rm H\alpha} = 4\pi D^2 F_{\rm H\alpha}$, where $D$ is the kinematic 
 distance and $F_{\rm H\alpha}$ is the H$\alpha$ flux corrected for 
 interstellar extinction by Balmer decrement from \cite{cap00} and the
 Galactic extinction properties (\citealt{sav79}). 
 Col.(3): Number of Lyman continuum photons estimated from
 col.(2). Col.(4): Electron number density from \cite{deh00}. Col.(5):
 Observed Oxygen abundance from \cite{deh00}. Col.(6): Dust-to-gas mass
 ratio determined from col.(5) by using the $\cal D$--(O/H) relation of
 figure~2 (see also \citealt{hir9a,hir9b}). Col.(7): Estimated $f$
 parameter by using equations~(7), and (10).}

\end{deluxetable}

\clearpage

\begin{deluxetable}{lcclccc}
\footnotesize
\tablecaption{H$\alpha$ -- IR correlation of the Galactic H {\sc ii}
 regions}
\tablewidth{0pt}
\tablehead{\colhead{Object} & \colhead{$E(B-V)$} 
& \colhead{$F_{\rm H\alpha}$} & \colhead{{\it IRAS} source} 
& \colhead{$F_{IRAS}$} & \colhead{$\epsilon$} & \colhead{$f$}\\
\colhead{} & \colhead{(mag)} 
& \colhead{($10^{-10}$ erg s$^{-1}$ cm$^{-2}$)} & \colhead{} 
& \colhead{($10^{-8}$ erg s$^{-1}$ cm$^{-2}$)} 
& \colhead{} & \colhead{}\\ 
\colhead{(1)} & \colhead{(2)} & \colhead{(3)} & \colhead{(4)} &
 \colhead{(5)} & \colhead{(6)} & \colhead{(7)}\\}

\startdata
S83     & 1.75 & 6.88 & 19223+2041 & 0.98 & 1    & 1.35 \\
S93     & 1.76 & 12.6 & 19529+2704 & 6.91 & 1    & 0.49 \\
S100    & 1.31 & 19.4 & 19598+3324 & 50.8 & 1    & 0.11 \\
S101    & 0.56 & 6.07 & 19579+3509 & 0.22 & 0.98 & 2.49 \\
S127    & 1.65 & 2.77 & 21270+5423 & 1.29 & 1    & 0.56 \\
S128    & 1.83 & 6.31 & 21306+5540 & 3.53 & 1    & 0.48 \\
S138    & 1.38 & 2.80 & 22308+5812 & 4.15 & 1    & 0.20 \\
S146    & 1.57 & 6.16 & 22475+5939 & 4.05 & 1    & 0.42 \\
S148    & 0.99 & 4.09 & 22542+5815 & 2.73 & 1    & 0.41 \\
S152    & 1.10 & 9.28 & 22566+5830 & 6.68 & 1    & 0.38 \\
S153    & 0.68 & 2.45 & 22571+5828 & 2.84 & 0.99 & 0.25 \\
S156    & 1.12 & 16.5 & 23030+5958 & 7.34 & 1    & 0.58 \\
S168    & 0.99 & 8.95 & 23504+6012 & 1.01 & 1    & 1.55 \\
S206    & 1.36 & 52.1 & 03595+5110 & 2.44 & 1    & 2.32 \\
S211    & 1.61 & 5.08 & 04329+5045 & 1.29 & 1    & 0.91 \\
S212    & 0.92 & 13.8 & 04368+5021 & 0.73 & 1    & 2.22 \\
S217    & 0.73 & 2.72 & 04551+4755 & 0.50 & 1    & 1.14 \\
S252    & 0.33 & 7.90 & 06066+2029 & 3.82 & 0.90 & 0.51 \\
\enddata

\tablecomments{Col.(2): Color excess determined from the Balmer
 decrement. Col.(3): H$\alpha$ fluxes corrected for interstellar
 extinction of col.(2) by the Galactic extinction properties
 (\citealt{sav79}). Col.(5): {\it IRAS} 40--120 $\mu$m fluxes derived by 
 $F_{IRAS}=1.26\times10^{-11}(2.56F_{60}/{\rm Jy}+F_{100}/{\rm Jy})$.
 Col.(6): Estimated $\epsilon$ by using equation~(\ref{eq42}). Col.(7):
 Estimated $f$ by using equation~(\ref{eq41}) and col.(6).}

\end{deluxetable}

\clearpage
\begin{deluxetable}{lcccccccc}
\footnotesize
\tablecaption{$\hat f$ of Seven Nearby Spiral Galaxies}
\tablewidth{0pt}
\tablehead{\colhead{Object} & \colhead{Distance} 
& \colhead{$F_{100}$} & \colhead{$F_{200}$} & \colhead{$T_{\rm d}$} 
& \colhead{$\log M_{\rm d}$} & \colhead{$\log M_{\rm HI + H_2}$} 
& \colhead{$\cal D$} & \colhead{$\hat{f}$}\\
\colhead{} & \colhead{Mpc} & \colhead{(Jy)} 
& \colhead{(Jy)} & \colhead{(K)} & \colhead{($M_\sun$)} 
& \colhead{($M_\sun$)} & \colhead{($10^{-3}$)} & \colhead{}\\ 
\colhead{(1)} & \colhead{(2)} & \colhead{(3)} & \colhead{(4)} &
 \colhead{(5)} & \colhead{(6)} & \colhead{(7)} & \colhead{(8)}
& \colhead{(9)}\\}

\startdata
NGC 134     & 21.1 & 67.0 & 127 & 21 & 8.18 & 10.40 & 5.9 & 0.21 \\ 
NGC 628     & 8.8  & 67.5 & 209 & 19 & 7.72 & 10.04 & 4.8 & 0.29 \\
NGC 660     & 11.3 & 120  & 158 & 24 & 7.51 & 10.04 & 2.9 & 0.48 \\ 
NGC 5194    & 6.2  & 303  & 533 & 22 & 7.62 & 9.88  & 5.5 & 0.24 \\ 
NGC 5236    & 6.9  & 624  & 889 & 23 & 7.90 & 10.54 & 2.3 & 0.56 \\ 
NGC 6946    & 10.1 & 338  & 743 & 20 & 8.38 & 10.48 & 8.0 & 0.12 \\ 
NGC 7331    & 10.9 & 120  & 243 & 21 & 7.85 & 10.00 & 7.0 & 0.16 \\ 
\enddata

\tablecomments{
Col.(1): Object name. 
Col.(2): Distance from our Galaxy in Mpc from \cite{alt98}.
Cols.(3) -- (4): Integrated flux density of {\it IRAS}-HIRES image at
 100, 200 $\mu$m, respectively, from \cite{alt98}.
Col.(5): Dust temperature estimated from $F_{100}/F_{200}$ and the
 modified black-body spectrum with the spectral index of 1.
Col.(6): Dust mass estimated from $F_{100}$ and $T_{\rm d}$ via
 equation~(\ref{eq50}). 
Col.(7): Total (H {\sc i} + H$_2$) gas mass. The data of six galaxies
 except NGC 134 are taken from \cite{dev90}. H {\sc i} and H$_2$ data of 
 NGC 134 are taken from \cite{mat96} and \cite{elf96}, respectively. In
 the calculation of mass, the same CO--H$_2$ conversion factor, 
$X_{\rm CO}=2.8\times10^{20}$ [(H$_2$ cm$^{-2}$)/(K km s$^{-1}$)], and the
 distances in col.(2) are used for all the sample galaxies.
Col.(8): Dust-to-gas mass ratio.
Col.(9): $\hat{f}$ estimated $\cal D$ and equations~(\ref{eq31}) and
 (\ref{eq30}) with $x=2$.}

\end{deluxetable}

\clearpage
\begin{figure}
\plotone{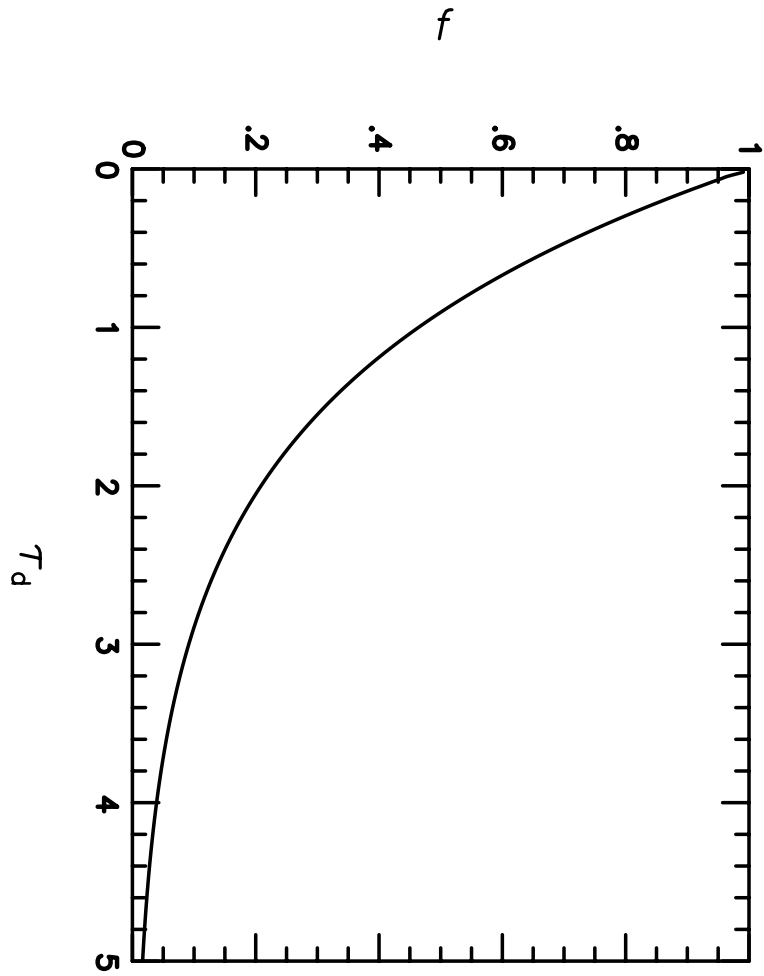}
\figcaption{$f$ vs. dust optical depth over actual ionized radius,
$\tau_{\rm d}$.}
\end{figure}

\clearpage
\begin{figure}
\plotone{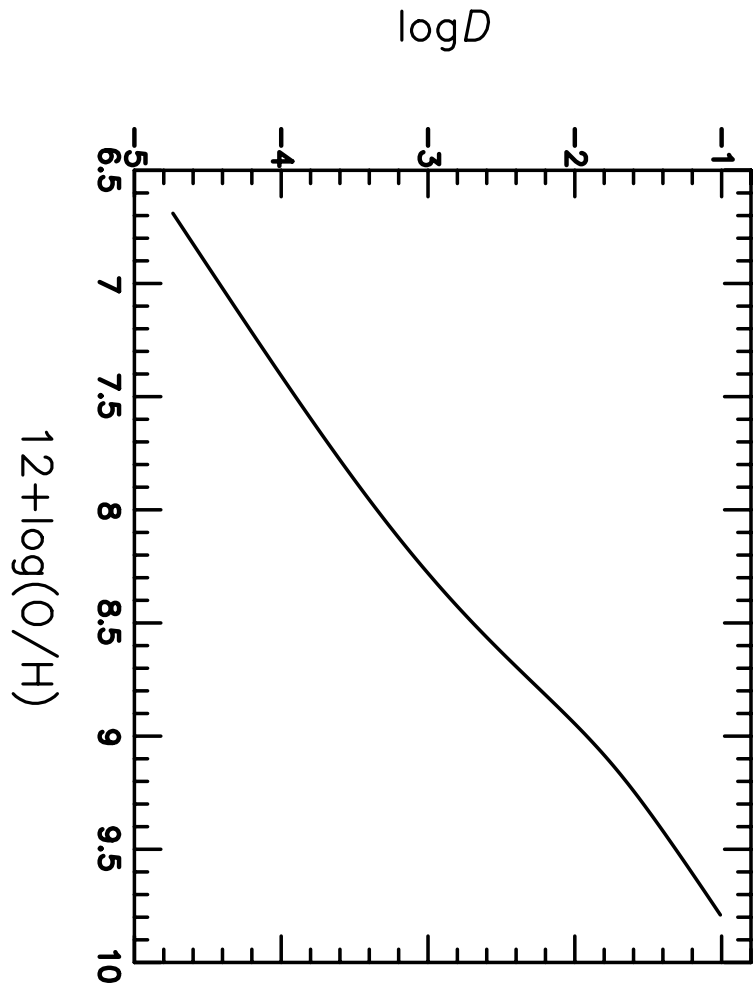}
\figcaption{Global relation between $\cal D$ and Oxygen abundance
proposed by \cite{hir9a, hir9b}.}
\end{figure}

\clearpage
\begin{figure}
\epsscale{0.9}
\plottwo{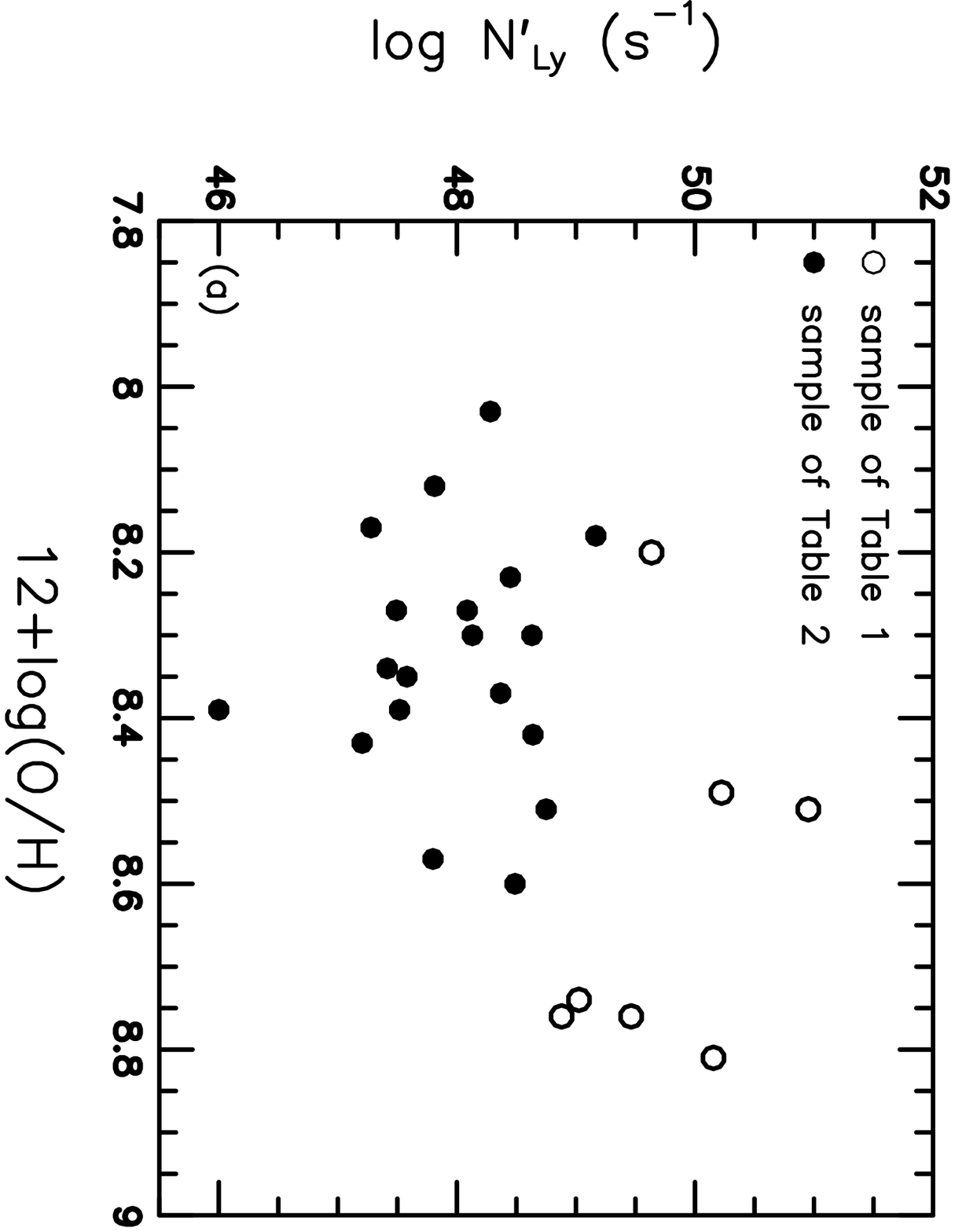}{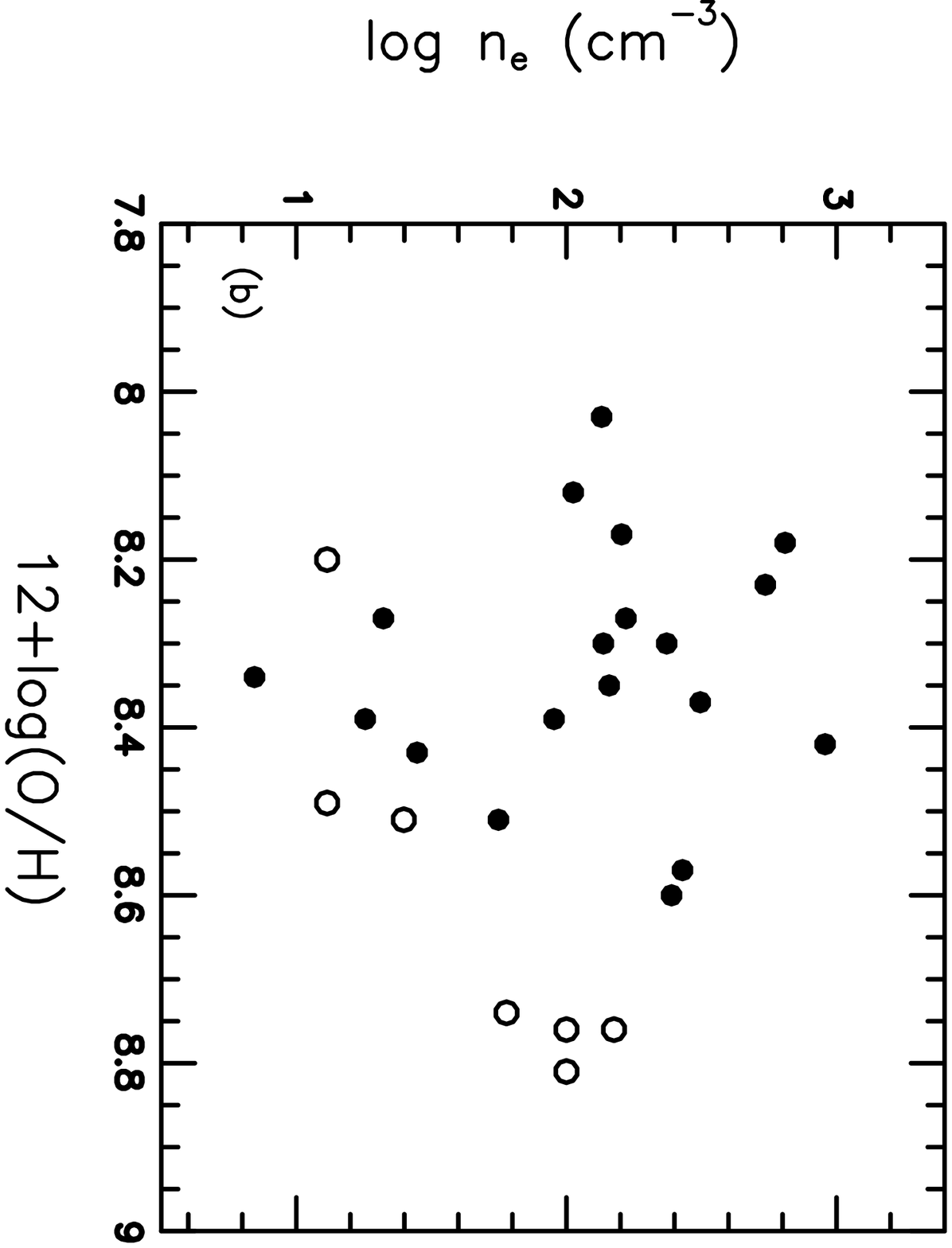}
\end{figure}

\begin{figure}
\plottwo{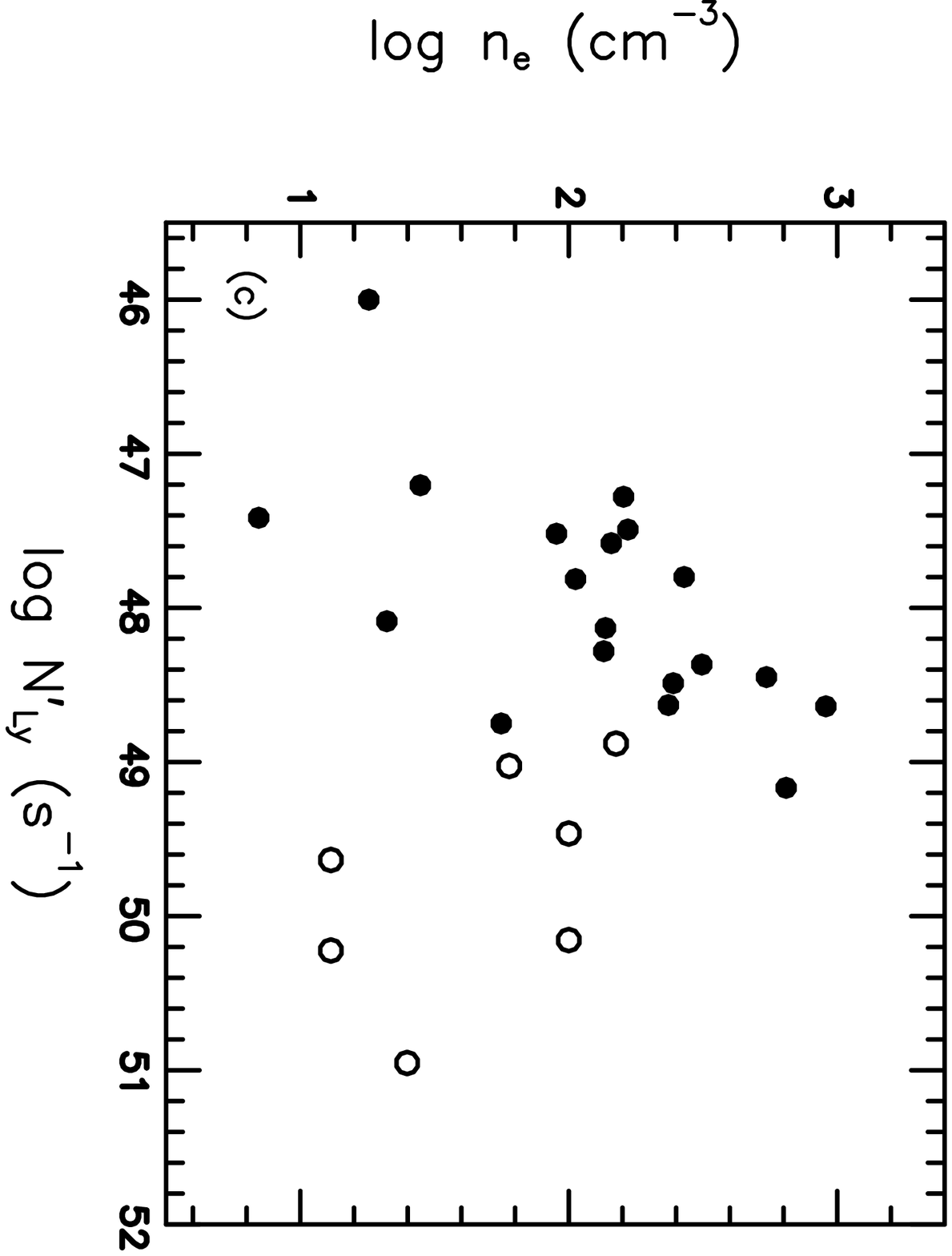}{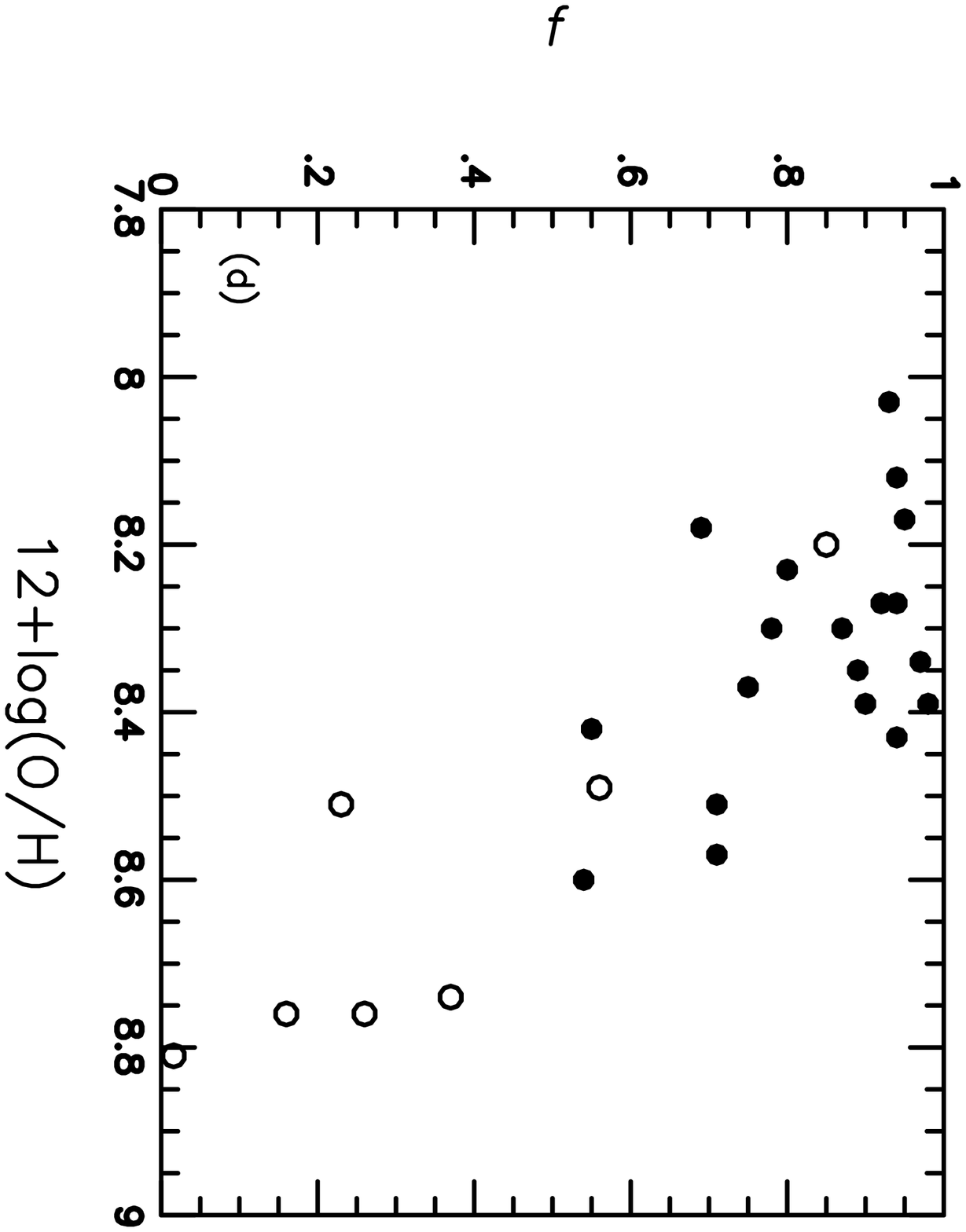}
\end{figure}

\begin{figure}
\plottwo{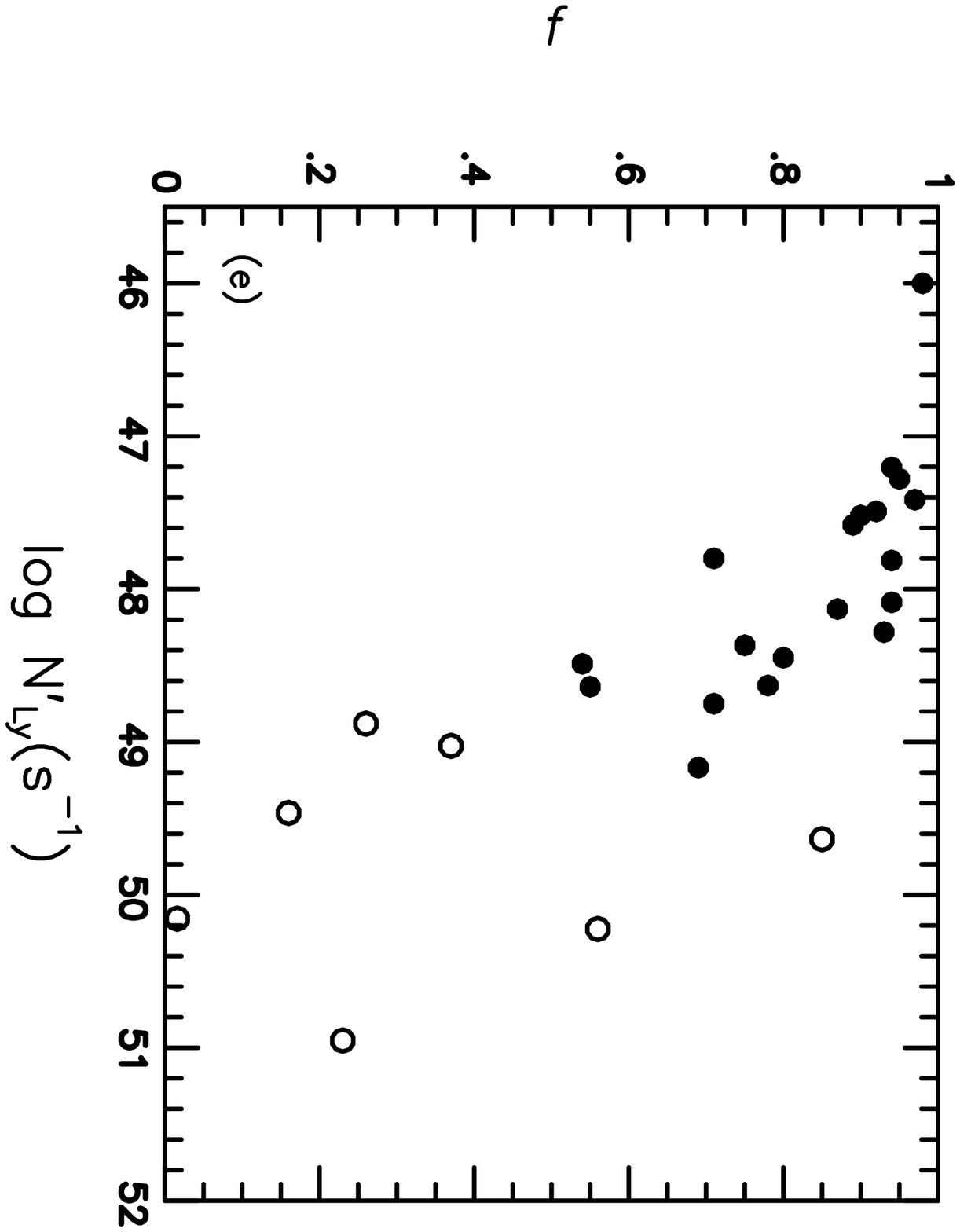}{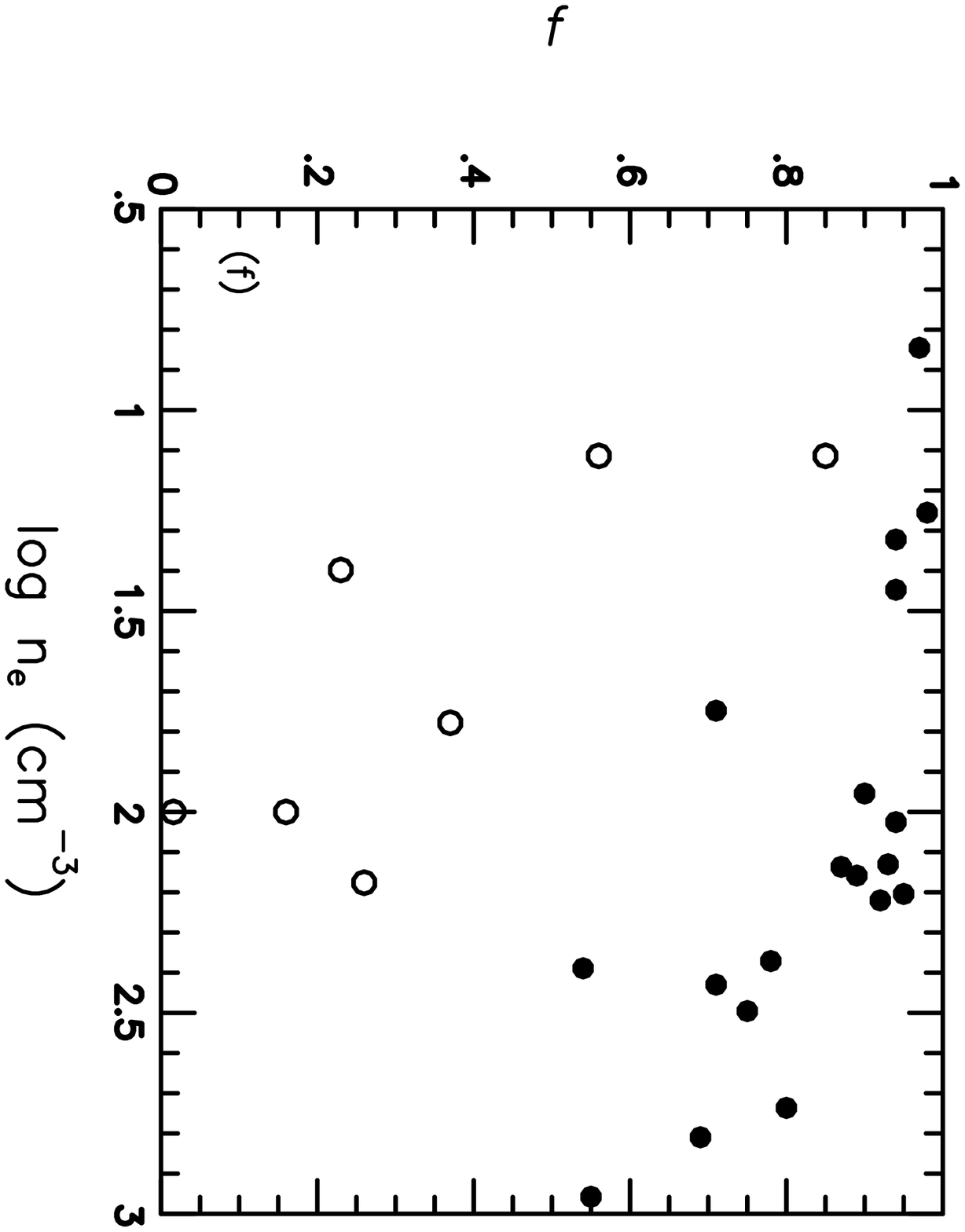}
\figcaption{Various correlations among various properties of our sample
of the Galactic H {\sc ii} regions. Open symbols are the samples in
Table~1, and Filled symbols are the samples in Table~2. (a) metallicity
vs. observed Lyman continuum number per unit time, $N'_{\rm Ly}$. (b)
metallicity vs. electron number density, $n_{\rm e}$. (c) $N'_{\rm Ly}$
vs. $n_{\rm e}$. (d) Estimated $f$ vs. metallicity. (e) $f$ vs. $N'_{\rm
Ly}$. (f) $f$ vs. $n_{\rm e}$.}
\end{figure}

\clearpage
\begin{figure}
\plotone{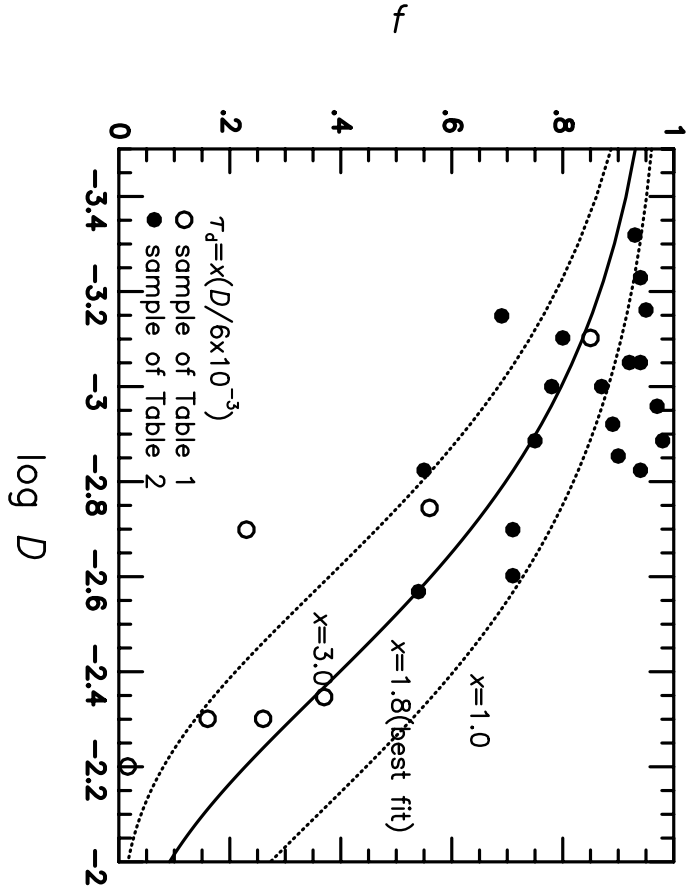}
\figcaption{Estimated $f$ vs. dust-to-gas mass ratio, $\cal D$.
Open symbols are the representative giant H {\sc ii} regions in
Table~1, and filled symbols are H {\sc ii} regions in Table~2. This
figure is basically the same as figure~3 (d).
The solid line denotes the best fit, $x=1.8$ in
equation~(\ref{eq30}). Two dotted lines mean $x=1.0$ and 3.0.}
\end{figure}

\clearpage
\begin{figure}
\plotone{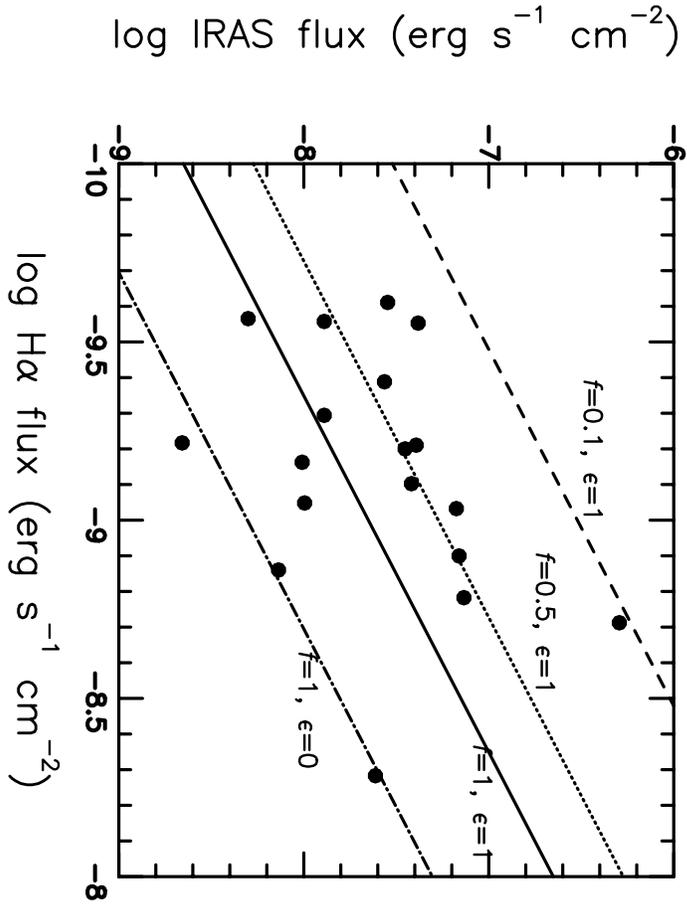}
\figcaption{{\it IRAS} flux vs.\ H$\alpha$ flux of some Galactic H {\sc ii}
regions. The plotted points are the sample H {\sc ii} regions of
Table~3. The lines are the $F_{IRAS}$--$F_{\rm H\alpha}$ relation
expected by equation~(\ref{eq41}). The solid, dotted, and dashed lines
mean $f = 1$, 0.5, and 0.1, respectively, adopting $\epsilon = 1$. We
set $f=1$, $\epsilon=0$ for the dash-dotted line.}
\end{figure}

\clearpage
\begin{figure}
\plotone{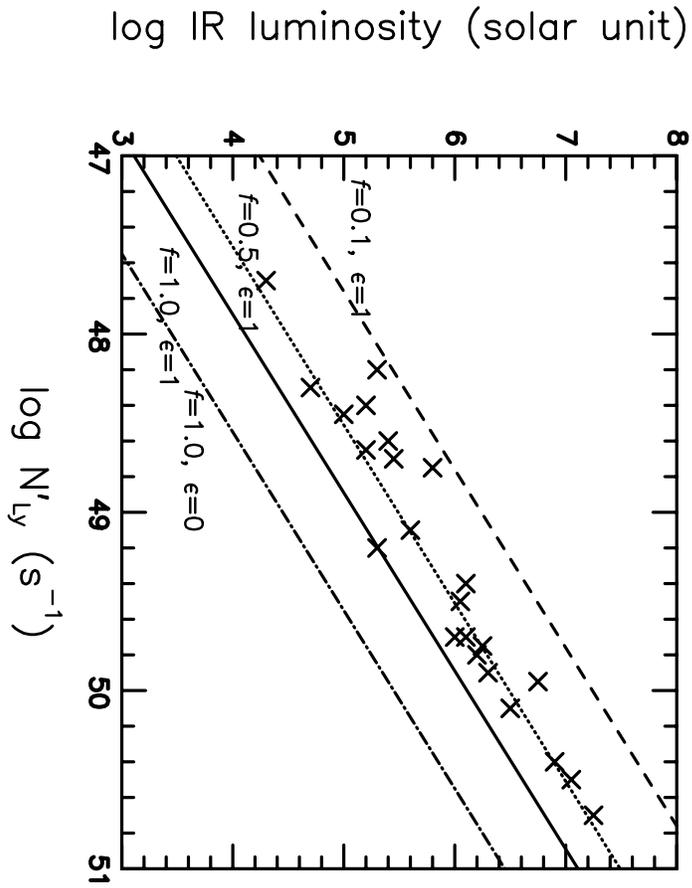}
\figcaption{Relation between IR luminosity and number of Lyman
continuum photons. The crosses are observational
data for H {\sc ii} regions reported by \cite{wyn74}. The parameter
sets of lines are the same as those of figure~5.}
\end{figure}

\end{document}